# Adiabatic transverse thermoelectric conversion enhanced by heat current manipulation in artificially tilted multilayers

Fuyuki Ando,[1,*] Takamasa Hirai,[1] Hiroto Adachi,[2] and Ken-ichi Uchida[1,3]

[1]*Research Center for Magnetic and Spintronic Materials, National Institute for Materials Science, 1-2-1 Sengen, Tsukuba, Ibaraki 305-0047, Japan*
[2]*Research Institute for Interdisciplinary Science, Okayama University, 3-1-1 Tsushimanaka, Kita-ku, Okayama, Okayama 700-8530, Japan*
[3]*Department of Advanced Materials Science, Graduate School of Frontier Sciences, The University of Tokyo, 5-1-5 Kashiwanoha, Kashiwa, Chiba 277-8561, Japan*

We phenomenologically formulate and experimentally observe an adiabatic transverse thermoelectric conversion enhanced by a heat current re-orientation in artificially tilted multilayers (ATMLs). By alternately stacking two materials with different thermal conductivities and rotating its multilayered structure with respect to a longitudinal temperature gradient, off-diagonal components in the thermal conductivity tensor are induced. This off-diagonal thermal conduction (ODTC) generates a finite transverse temperature gradient and Seebeck-effect-induced thermopower in the adiabatic condition, which is superposed on the isothermal transverse thermopower driven by the off-diagonal Seebeck effect (ODSE). In this study, we calculate and observe the two-dimensional temperature distribution and the resultant transverse thermopower in ATMLs comprising thermoelectric $Co_2MnGa$ Heusler alloys and $Bi_{2-a}Sb_aTe_3$ compounds. By changing the tilt angle from 0° to 90°, the transverse temperature gradient obviously appeared in the middle angles and the transverse thermopower increases up to -116.1 μV/K in $Co_2MnGa/Bi_{0.2}Sb_{1.8}Te_3$-based ATML at the tilt angle of 45° whereas the isothermal contribution is estimated to be -82.6 μV/K from the analytical calculation. This hybrid action derived from ODTC results in the significant variation of the maximum reduced efficiency for transverse thermoelectric conversion from 3.1% in the isothermal limit to 8.1% in the adiabatic limit.

## I. INTRODUCTION

A transverse thermoelectric effect in solids, which mutually converts charge and heat currents in the orthogonal directions, has attracted much attention in terms of fundamental physics, materials science, and thermoelectric applications [1–4]. This unique geometry offers a practical advantage, that is, simplification of a thermoelectric device architecture only using a single material, whereas classical thermoelectric devices utilizing the Seebeck effect consist of multiple thermoelectric materials and electrodes connecting in series and forming a π-shaped structure. The most representative transverse thermoelectric effect is the Nernst effect [1]. Recently, associated with the development of spin caloritronics [4–7] and topological materials science [8–10], the studies on both the ordinary and anomalous Nernst effects have rapidly advanced and achieved the superior transverse thermoelectric performance than ever before owing to the non-trivial band topology [11–20] or extrinsic mechanisms [21–23]. Meanwhile, ODSE as the other principle has been intensively studied for wide variety of materials having microscale anisotropy, i.e., the goniopolar materials [24–28] and ($p \times n$)-type thermoelectric superlattices [29,30], and having macroscale anisotropy, i.e., ATMLs [31–35]. The ODSE materials exhibit higher performances than those by the Nernst effects. Also, the hybridization of the multiple principles in a single material also provides a key for the further advance of materials science for transverse thermoelectrics [36,37]. These recent progresses are expected to develop next core technologies for energy harvesting and thermal managements.

In addition to the insight into materials characteristics, the thermal boundary conditions, i.e., isothermal or adiabatic, also have a significant influence on the transverse thermoelectric performance [26]. The standard definition of the figure of merit for transverse thermoelectric conversion $z_{xy}T$ is given as [38–41]:

*Contact author: ANDO.Fuyuki@nims.go.jp

$$z_{xy}T = \frac{S_{xy}^2}{\rho_{xx}\kappa_{yy}}T. \tag{1}$$

where $S_{xy} \equiv E_x/\nabla_y T|_{\nabla_x T=0}$ is the isothermal transverse thermopower by the ratio of the applied temperature gradient in y-axis ($\nabla_y T$) and generated electric field in x-axis ($E_x$) under the temperature gradient in x-axis $\nabla_x T = 0$ as shown in Fig. 1(a). Also, $\rho_{xx} \equiv E_x/j_{c,x}|_{\nabla_x T=0}$ and $\kappa_{yy} \equiv -j_{q,y}/\nabla_y T|_{\nabla_x T=0}$ are respectively the isothermal electrical resistivity and thermal conductivity with $j_{c,x}$ and $j_{q,y}$ being the charge current density in x-axis and heat current density in y-axis. Delves [42,43] and Horst [44,45] pointed out that when the isothermal condition of $\nabla_x T = 0$ is replaced by the adiabatic condition of $j_{q,x} = 0$, $z_{xy}T$ changes to an adiabatic figure of merit $z_{xy}^*T$ due to the appearance of finite $\nabla_x T$ and modification to the adiabatic transverse thermopower $S_{xy}^*$, electrical resistivity $\rho_{xx}^*$, and thermal conductivity $\kappa_{yy}^*$. Note that the adiabatic physical quantities are highlighted by the superscript * in this work. $\nabla_x T$ typically originates from ODTC of the applied $j_{q,y}$, such as the thermal Hall effect (i.e., Righi-Leduc effect) or anisotropy of thermal conductivity. However, the adiabatic transverse thermoelectric properties modulated by ODTC-induced $\nabla_x T$ has generally been considered for the cooling applications [12,42,44,46,47] or treated just as a correction term to explain why magneto-thermoelectric coefficients differ in the same materials, e.g., the difference in thermal boundary conditions due to the aspect ratio [19,26,43,45]. Thus, there has been no attempt to constructively utilize ODTC to improve transverse thermoelectric performance.

Materials which actively control the heat current direction have been developed to form unique thermal circuits harvesting waste heat, which are often called thermal metamaterials [48–50]. Figure 1(b) shows a schematic of ATML consisting of two materials with different thermal conductivities $\kappa$, where the effective thermal conductivities in parallel and perpendicular to the stacking plane ($\kappa_\parallel$ and $\kappa_\perp$) are different from each other. Owing to this structure-induced anisotropic thermal conductivity, the heat current bends from y-axis when $\nabla_y T$ is applied oblique to the stacking plane. Here, the off-diagonal component of the thermal conductivity tensor $\kappa_{xy} = -j_{q,x}/\nabla_y T$ in ATMLs is expressed by the analytical matrix calculation in Appendix A with Refs. [51–53] as

$$\kappa_{xy} = (\kappa_\parallel - \kappa_\perp)\sin\theta\cos\theta, \tag{2}$$

where $\theta$ denotes the tilt angle of the stacking plane to x-axis. The $(\kappa_\parallel - \kappa_\perp)$ term increases as the difference in $\kappa$ between the constituent materials increases. Then, the thermally adiabatic limit and open circuit condition [$j_{q,x} = j_{c,x} = 0$ in Eq. (A1)] gives a relationship between $\kappa_{xy}$ and $\nabla_x T$ as follows:

$$\nabla_x T = -\frac{\kappa_{xy}}{\kappa_{xx}}\nabla_y T. \tag{3}$$

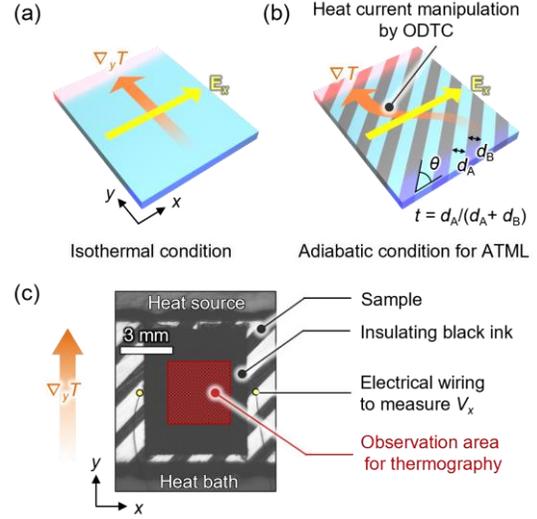

FIG 1. Schematic of transverse thermoelectric conversion in (a) isothermal condition and (b) adiabatic condition for an anisotropic material, such as ATML. By rotating the multilayer by $\theta$ to the x-axis, ODTC is induced so that the bent heat current direction modulates $E_x$. (c) An infrared image of CMG/BT-based ATML with $\theta = 45°$ placed on the measurement setup to characterize two-dimensional temperature distribution and $V_x$.

The previous studies experimentally observed sizable bending angles up to 26° in copper/stainless use steel-based ATMLs, which corresponds to $\nabla_x T/\nabla_y T \sim -0.5$ owing to the large $-\kappa_{xy}/\kappa_{xx}$ [48,52]. Thus, there is abundant room for utilizing the bending heat current for transverse thermoelectrics by hybridizing transverse and longitudinal thermoelectric effects driven by $\nabla_y T$ and $\nabla_x T$.

In this study, we phenomenologically formulate and experimentally demonstrate the enhancement of $S_{xy}^*$ and $z_{xy}^*T$ by combining ODSE induced by $\nabla_y T$ and the

*Contact author: ANDO.Fuyuki@nims.go.jp

Seebeck effect induced by $\nabla_x T$ in ATMLs. We synthesized the ATML slabs comprising thermoelectric Co$_2$MnGa Heusler alloys and Bi$_{2-a}$Sb$_a$Te$_3$ compounds with various $\theta$ from 0° to 90°. Through the observation of two-dimensional temperature distribution using an infrared camera while applying $\nabla_y T$, the $\theta$ dependence of ODTC-induced $\nabla_x T/\nabla_y T$ was characterized and confirmed the consistency with the analytically calculated $-\kappa_{xy}/\kappa_{xx}$. As the result of the heat current re-orientation, the higher transverse thermopower of -116.1 μV/K than $S_{xy}$ of -82.6 μV/K obtained by the conventional matrix calculation was observed in Co$_2$MnGa/Bi$_{0.2}$Sb$_{1.8}$Te$_3$-based ATML. Owing to the ODTC-induced enhancement of transverse thermopower, the analytically calculated $z_{xy}^*T$ reaches 0.28 in maximum at room temperature whereas the maximum $z_{xy}T$ is calculated to be 0.13, which corresponds to the 165% improvement for the transverse thermoelectric conversion efficiency. The heat current manipulation in the adiabatic condition thus provides a distinct strategy on the material developments and device designs for transverse thermoelectrics.

## II. METHODS

To obtain a superior ODSE-induced transverse thermopower, Co$_2$MnGa (CMG), Bi$_{0.2}$Sb$_{1.8}$Te$_3$ (BST), and Bi$_2$Te$_3$ (BT) were selected as the constituent materials for ATMLs. The synthesis processes and measurement methods for thermoelectric transport properties for each sintered body are detailly described in Ref. [37]. Table I shows the summary of the measurement results: the Seebeck coefficient $S_{SE}$, electrical conductivity $\sigma$, and thermal conductivity $\kappa$ for CMG, BST, and BT. It is found that the transport properties, especially $\kappa$, are greatly different between CMG and BST (BT), which is favorable to obtain both large ODSE and ODTC in ATMLs.

TABLE I. Thermoelectric transport properties for CMG, BST, and BT from Ref. [37].

|  | CMG | BST | BT |
|---|---|---|---|
| $S_{SE}$ (μV/K) | -32.1 | 170.6 | -110.3 |
| $\sigma$ (×10$^5$ S/m) | 7.99 | 1.28 | 1.83 |
| $\kappa$ (W/m·K) | 18.7 | 1.2 | 1.5 |

*Contact author: ANDO.Fuyuki@nims.go.jp

The CMG/BST- and CMG/BT-based ATML slabs with various $\theta$ values were prepared by the spark plasma sintering method. The sintered CMG with a diameter of 20 mm was sliced into many disks with a thickness of 1 mm using a diamond wire saw. The CMG disks and BST (BT) powders were alternately filled into a graphite die and sintered at 450°C with a uniaxial pressure of 30 MPa and soaking time of 60 min under high vacuum. The averaged thickness of BST (BT) was 0.7 mm, which corresponds to the thickness ratio of CMG $t = d_{CMG}/d_{CMG}+d_{BST(BT)} = 0.59$ with $d$ being the thickness of each layer. The sintered CMG/BST and CMG/BT multilayers were cut into rectangular slabs with a size of 10 × 8 × 1 mm$^3$ and $\theta$ = 0, 15, 30, 45, 60, 75, and 90°.

The two-dimensional temperature distribution and transverse thermopower for CMG/BST- and CMG/BT-based ATMLs were measured using an infrared camera and homemade thermopower measurement setup. Figure 1(c) shows an infrared image of CMG/BT-based ATML with $\theta$ = 45° placed on the sample holder. The sample was bridged between two anodized Al blocks in the y-direction with a distance of ~6 mm, one of which acted as a heat source by applying a charge current to connected chip heaters and the other as a heat sink to generate $\nabla_y T$. Thus, most part of the ATML sample floated in the air keeping a region of interest close to the adiabatic condition. The central part of 10 × 8 mm$^2$ surface was covered with an insulating black ink having a high emissivity over 0.94. To measure a transverse thermoelectric voltage, two Al-1%Si wires were directly attached on the 10 × 8 mm$^2$ surface with a distance of ~6.5 mm in the x-direction by using a wire bonder. The observation area for thermography using an infrared camera is defined as a 3.6 × 3.6 mm$^2$ square centered on the midpoint of the two Al-1%Si electrodes.

## III. RESULTS

### A. Off-diagonal thermal conduction

We begin with the analytical calculation of the thermal conductivity tensor for CMG/BST- and CMG/BT-based ATMLs following Appendix A and Refs. [51–53]. The measured $\kappa$ in Table I and $t$ are substituted into Eqs. (A1) and (A2) to calculate $\kappa_{ij}$. According to the equations, $t$ and $\theta$ are variables for $\kappa_{ij}$ as well as $\kappa$ of CMG, BST, and BT. Figures 2(a)−(c) shows the contour plots of $\kappa_{xx}$, $\kappa_{xy}$, and $-\kappa_{xy}/\kappa_{xx}$ as functions of $t$ and $\theta$ for CMG/BST-based ATML. Figure 2(a) shows that $\kappa_{xx}$ monotonically increases as $t$ ($\theta$) increases (decreases) due to the enhanced contribution of CMG with high $\kappa$.

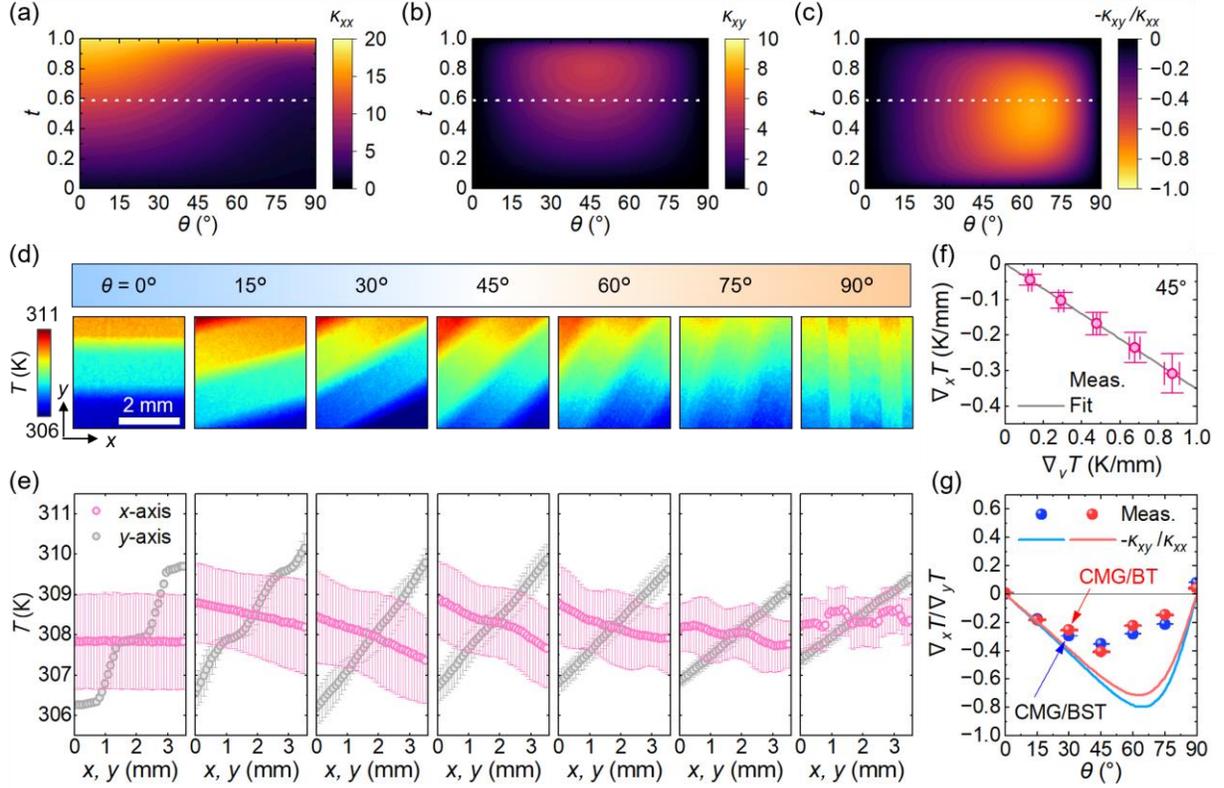

FIG 2. (a)-(c) Contour plots of (a) $\kappa_{xx}$, (b) $\kappa_{xy}$, and (c) $-\kappa_{xy}/\kappa_{xx}$ obtained by the matrix calculation in Appendix A. White dotted lines indicate $t = 0.59$. (d) Thermal images of CMG/BST-based ATMLs with various $\theta$ values under the application of $\nabla_y T$. (e) The $x$- and $y$-axes line profiles of the average temperature signals in the $y$- and $x$-directions. (f) The plot of effective $\nabla_x T$ and $\nabla_y T$ for CMG/BST-based ATML with $\theta = 45°$ under various temperature differences between the heat source and sink. (g) The $\theta$ dependence of $\nabla_x T/\nabla_y T$ for CMG/BT- and CMG/BST-based ATMLs with the calculated curves of $-\kappa_{xy}/\kappa_{xx}$ at $t = 0.59$.

Meanwhile, $\kappa_{xy}$ in Fig. 2(b) shows a different trend maximizing at 45° due to the $\sin\theta \cos\theta$ contribution. As a result, $-\kappa_{xy}/\kappa_{xx}$ in Fig. 2(c) shows a unique $\theta$ dependence which minimizes at 64° and relatively moderate $t$ dependence. In the case of CMG/BT-based ATML, $-\kappa_{xy}/\kappa_{xx}$ minimizes at 62°, suggesting that the optimum $t$ and $\theta$ to maximize ODTC depend on the balance of $\kappa$ for constituent materials.

To experimentally confirm the heat current manipulation by ODTC, we observed and analyzed the two-dimensional temperature distribution for our ATMLs. Figure 2(d) shows thermal images in the area indicated by Fig. 1(c) under the application of $\nabla_y T$ for CMG/BST-based ATMLs. Here, the input heater power was tuned so that the temperature deference between the heat source and sink was stabilized at 10.0 ± 0.1 K. All the thermal images represent layered patterns with different angles corresponding to $\theta$, which reflects to the difference in $\kappa$ between CMG and BST. The temperature drastically changes in the BST layers due to the lower $\kappa$. To quantitatively discuss ODTC-induced $\nabla_x T$ and $\nabla_y T$ in our ATMLs, we plot the $x$- and $y$-axes position dependence of the temperature signals averaged in the $y$- and $x$-directions as shown in Fig. 2(e). The coordinate origin is defined at the lower left corner of the thermal images. Importantly, the negative slope in the $x$-axis plot is clearly observed only in the middle $\theta$ values, whereas those are almost zero at $\theta = 0°$ and 90°. This characteristic is consistent with $\kappa_{xy} \propto \sin\theta \cos\theta$ in Fig. 2(b). The $y$-axis plot shows the decrease in the positive slope as $\theta$ increases, which can result from the increase in $\kappa_{yy}$ of the ATML samples according to the matrix calculation in Appendix A.

We compare analytically calculated $-\kappa_{xy}/\kappa_{xx}$ and experimentally measured $\nabla_x T/\nabla_y T$ for CMG/BST- and CMG/BT-based ATMLs. Figure 2(f) shows the plot of effective $\nabla_x T$ and $\nabla_y T$ for CMG/BST-based

*Contact author: ANDO.Fuyuki@nims.go.jp

ATML with $\theta = 45°$ under the various temperature differences between the heat source and sink, where the $\nabla_x T$ and $\nabla_y T$ values are determined by the slope of linear fit in Fig. 2(e). Also, the $\nabla_x T/\nabla_y T$ values are evaluated by the slope of linear fit as shown in Fig. 2(f). The $\theta$ dependence of $\nabla_x T/\nabla_y T$ for all the ATML samples as well as the calculated curves for $-\kappa_{xy}/\kappa_{xx}$ at $t = 0.59$ are depicted in Fig. 2(g). As the measurement condition gets closer to the adiabatic (isothermal) limit, $\nabla_x T/\nabla_y T$ converges to $-\kappa_{xy}/\kappa_{xx}$ (zero). It is found that $\nabla_x T/\nabla_y T$ shows the similar negative trend as $-\kappa_{xy}/\kappa_{xx}$, suggesting that the measurement condition is relatively close to the adiabatic limit in this study. The $\nabla_x T/\nabla_y T$ values quantitatively agree with $-\kappa_{xy}/\kappa_{xx}$ especially at low $\theta$ region and reach the minimum value of ~0.4 at $\theta = 45°$ owing to the large difference in $\kappa$ in Table I, which is expected to have a significant impact on $S^*_{xy}$. The reason for the deviation at higher $\theta$ is mentioned in Discussion section. Hereby, we demonstrated the manipulation of the heat current direction and appearance of $\nabla_x T$ through ODTC for ATMLs in the adiabatic condition.

## B. Adiabatic transverse thermopower

We phenomenologically formulate the adiabatic transverse thermopower $S^*_{xy}$ by ODSE and the ODTC-induced Seebeck effect for anisotropic materials such as ATMLs. The two-dimensional thermoelectric tensor in the $x$-$y$ plane for the heat current **q** and electric field **E** is introduced as follows [42–45]:

$$\begin{pmatrix} j_{q,x} \\ j_{q,y} \\ E_x \\ E_y \end{pmatrix} = \begin{pmatrix} S_{xx}T & S_{xy}T & -\kappa_{xx} & -\kappa_{xy} \\ S_{yx}T & S_{yy}T & -\kappa_{yx} & -\kappa_{yy} \\ \rho_{xx} & \rho_{xy} & S_{xx} & S_{xy} \\ \rho_{yx} & \rho_{yy} & S_{yx} & S_{yy} \end{pmatrix} \begin{pmatrix} j_{c,x} \\ j_{c,y} \\ \nabla_x T \\ \nabla_y T \end{pmatrix}. \quad (4)$$

Here, we consider generation of $E_x$ by the application of $\nabla_y T$ under an open-circuit condition in the $y$-direction ($j_{c,y} = 0$). Then, the relevant linear-response equations can be written as

$$j_{q,x} = S_{xx} T j_{c,x} - \kappa_{xx} \nabla_x T - \kappa_{xy} \nabla_y T, \quad (5)$$

$$j_{q,y} = S_{yx} T j_{c,x} - \kappa_{yx} \nabla_x T - \kappa_{yy} \nabla_y T, \quad (6)$$

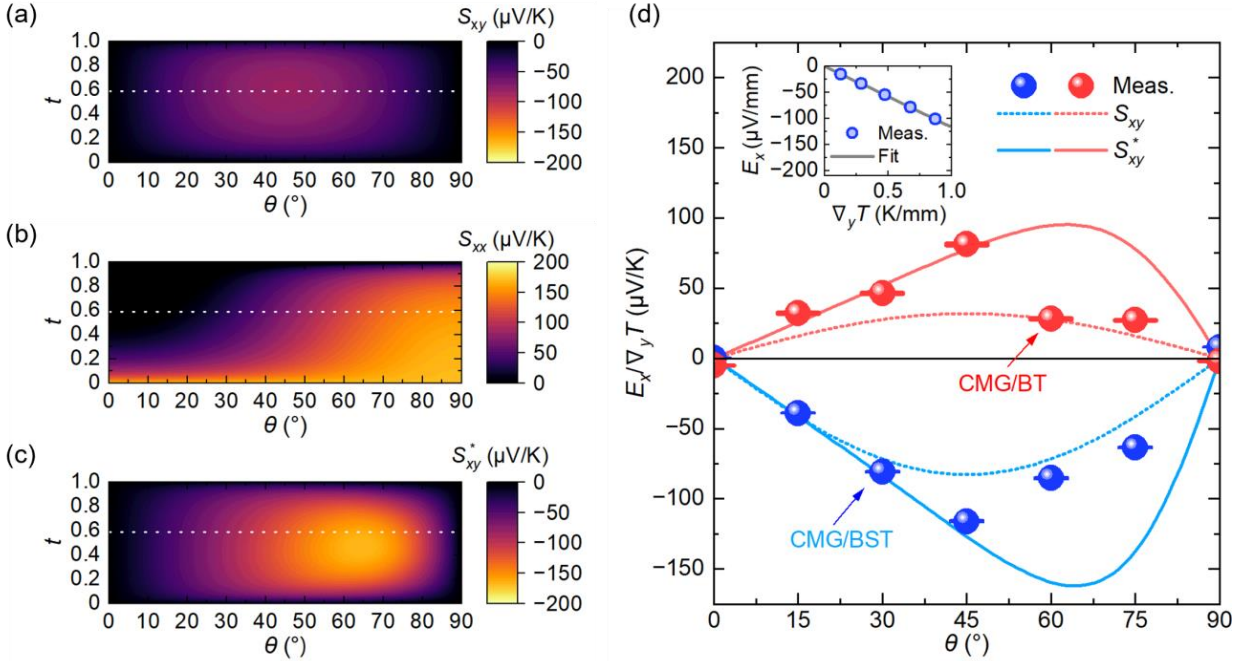

FIG 3. (a)-(c) Contour plots of (a) $S_{xy}$, (b) $S_{xx}$, and (c) $S^*_{xy}$ obtained by the matrix calculation for CMG/BST-based ATML. (d) The $\theta$ dependence of the measured $E_x/\nabla_y T$ and calculated $S_{xy}$ and $S^*_{xy}$ at $t = 0.59$ for CMG/BT- and CMG/BST-based ATMLs. The inset shows the $\nabla_y T$ dependence of $E_x$ for CMG/BST-based ATML with $\theta = 45°$.

*Contact author: ANDO.Fuyuki@nims.go.jp

$$E_x = \rho_{xx} j_{c,x} + S_{xx} \nabla_x T + S_{xy} \nabla_y T, \quad (7)$$

$$E_y = \rho_{yx} j_{c,x} + S_{yx} \nabla_x T + S_{yy} \nabla_y T. \quad (8)$$

Equations (5)–(8) in the adiabatic condition for the $x$-direction ($j_{q,x} = 0$) can be solved for $\nabla_y T$ and $E_x$ as

$$\nabla_y T = \frac{S_{yx}^*}{\kappa_{yy}^*} T j_{c,x} - \frac{1}{\kappa_{yy}^*} j_{q,y}, \quad (9)$$

$$E_x = \rho_{xx}^* j_{c,x} - \frac{S_{xy}^*}{\kappa_{yy}^*} j_{q,y}, \quad (10)$$

where the adiabatic transverse thermopower $S_{xy}^*$ and $S_{yx}^*$ are described by using the isothermal transverse thermopower $S_{xy}$ and $S_{yx}$ as follows:

$$S_{xy}^* = S_{xy} - \frac{\kappa_{xy}}{\kappa_{xx}} S_{xx}, \quad (11)$$

$$S_{yx}^* = S_{yx} - \frac{\kappa_{yx}}{\kappa_{xx}} S_{xx}, \quad (12)$$

where the first term represents the intrinsic ODSE and the second term the ODTC-induced Seebeck effect. In the case of ATMLs, both the Seebeck and thermal conductivity tensors are symmetric (see Appendix A), i.e., $S_{xy} = S_{yx}$, $\kappa_{xy} = \kappa_{yx}$, and hence $S_{xy}^* = S_{yx}^*$. Note that those for the Nernst and thermal Hall effects are antisymmetric due to Onsager's reciprocal relations, i.e., $S_{xy} = -S_{yx}$, $\kappa_{xy} = -\kappa_{yx}$, and hence $S_{xy}^* = -S_{yx}^*$.

Then, we analytically calculate $S_{xy}$ and $S_{xy}^*$. Figure 3(a) shows the contour plot of $S_{xy}$ as functions of $t$ and $\theta$ for CMG/BST-based ATML, calculated based on Eq. (11) and the thermoelectric transport tensors for ATMLs in Appendix A. The $S_{xy}$ value minimizes at 45° in a similar manner to $\kappa_{xy}$ in Fig. 2(b). On the other hand, because $S_{xx}$ and $-\kappa_{xy}/\kappa_{xx}$ have the different $t$ and $\theta$ dependences as shown in Figs. 3(b) and 2(c), $S_{xy}^*$ exhibits a distinct behavior with $S_{xy}$ [Fig. 3(c)]. Owing to the same sign of $S_{xy}$ and $-\kappa_{xy} S_{xx}/\kappa_{xx}$, the absolute value of $S_{xy}^*$ is greatly higher than that of $S_{xy}$ in this study: $S_{xy}^*$ reaches -167.7 µV/K at $t = 0.47$ and $\theta = 64°$, whereas $S_{xy}$ -82.7 µV/K at $t = 0.61$ and $\theta = 45°$. Thus, we naturally expect the improvement of the transverse thermopower through the contribution of ODTC.

Now, we are in the position to experimentally characterize the transverse thermopower $E_x/\nabla_y T$ for the comparison with $S_{xy}$ and $S_{xy}^*$. Figure 3(d) shows the measurement results for CMG/BST- and CMG/BT-based ATMLs. The $E_x/\nabla_y T$ values are determined by the slope of linear fit as shown in the inset of Fig. 3(d). Reflecting the opposite sign of $S_{SE}$ for BST and BT, CMG/BST- and CMG/BT-based ATMLs also show the opposite sign of $E_x/\nabla_y T$ (see Eqs. (A6), (A7), and (A10) of Appendix A for details). Interestingly, both the ATML systems exhibit higher $E_x/\nabla_y T$ than $S_{xy}$, which is the direct evidence of the constructive contribution of ODTC. The $E_x/\nabla_y T$ values quantitatively agree with $S_{xy}^*$ for $\theta \leq 45°$ but get closer to $S_{xy}$ with increasing $\theta$, whose trend is consistent with that of the measured $\nabla_x T/\nabla_y T$ in Fig.

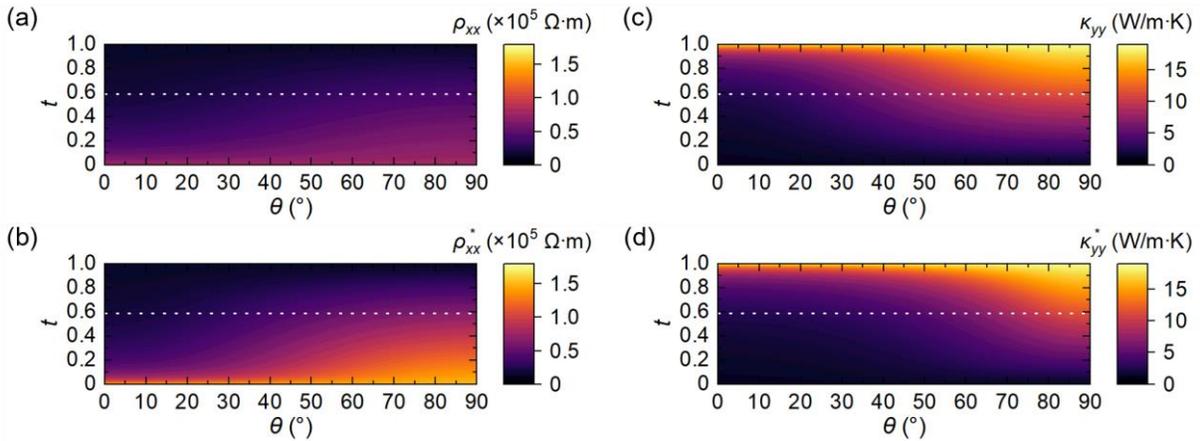

FIG 4. (a)-(d) Contour plots of (a) $\rho_{xx}$, (b) $\rho_{xx}^*$, (c) $\kappa_{xx}$, and (d) $\kappa_{xy}^*$ obtained by the matrix calculation for CMG/BST-based ATML.

*Contact author: ANDO.Fuyuki@nims.go.jp

2(g). Although the conventional matrix calculation suggests the optimum $S_{xy}$ of -82.6 µV/K at $\theta = 45°$ for CMG/BST-based ATML, $E_x/\nabla_y T$ measured in the adiabatic condition reaches the much higher value of -116.1 ± 1.2 µV/K, which leads to a significant enhancement of the transverse thermoelectric performance.

## C. Transverse thermoelectric performance

We phenomenologically formulate the adiabatic transport properties relevant to $z^*_{xy}T$ for anisotropic materials such as ATMLs. By solving Eqs. (5)−(8), $\rho^*_{xx}$ and $\kappa^*_{yy}$ defined by Eqs. (9) and (10) can be expressed as

$$\rho^*_{xx} = \rho_{xx} + \left(\frac{S^*_{xy}S^*_{yx}}{\kappa^*_{yy}} + \frac{S_{xx}^2}{\kappa_{xx}}\right)T, \quad (13)$$

$$\kappa^*_{yy} = \kappa_{yy} - \frac{\kappa_{xy}\kappa_{yx}}{\kappa_{xx}}. \quad (14)$$

Figures 4(a)−(d) show the analytically calculated $\rho_{xx}$, $\rho^*_{xx}$, $\kappa_{yy}$, and $\kappa^*_{yy}$ as functions of $t$ and $\theta$ for CMG/BST-based ATML, indicating the obviously different behaviors between the isothermal and adiabatic conditions as well as $S_{xy}$ and $S^*_{xy}$. Due to the symmetric relations of $S_{xy} = S_{yx}$ and $\kappa_{xy} = \kappa_{yx}$, $\rho^*_{xx}$ ($\kappa^*_{yy}$) necessarily exhibits a higher (lower) value than $\rho_{xx}$ ($\kappa_{yy}$). As a result, $z^*_{xy}T$ is expressed by the replacement of Eq. (1) by Eqs. (10), (12), and (13):

$$z^*_{xy}T = \frac{S^{*2}_{xy}}{\rho^*_{xx}\kappa^*_{yy}}T. \quad (15)$$

Note that by substituting the relations of $S_{xy} = -S_{yx}$, $\kappa_{xy} = -\kappa_{yx}$ due to Onsager's reciprocal theory to Eqs. (13)−(14), the adiabatic transport properties for the Nernst effects can be obtained, where $\kappa^*_{yy}$ necessarily exhibits a higher value than $\kappa_{yy}$ contrary to the case of ATMLs.

Figure 5(a)−(b) shows analytically calculated $z_{xy}T$ and $z^*_{xy}T$ values for CMG/BST-based ATML. In general, $S_{xy}$ ($\propto \sin\theta\cos\theta$) maximizes at $\theta = 45°$ and both $\rho_{xx}$ and $\kappa_{yy}$ monotonically decrease as $\theta$ decreases [Figs. 3(a), 4(a), and 4(c)]. Thus, the conventional calculation for ATMLs in the isothermal limit necessarily suggests the optimum $\theta$ lower than 45° to maximize $z_{xy}T$ [33–35,37]. In fact, Fig. 5(a) shows that $z_{xy}T$ for CMG/BST-based ATML maximizes to be 0.13 at $t = 0.56$ and $\theta = 30°$. On the other hand, the best $\theta$ to maximize $S^*_{xy}$ drastically

*Contact author: ANDO.Fuyuki@nims.go.jp

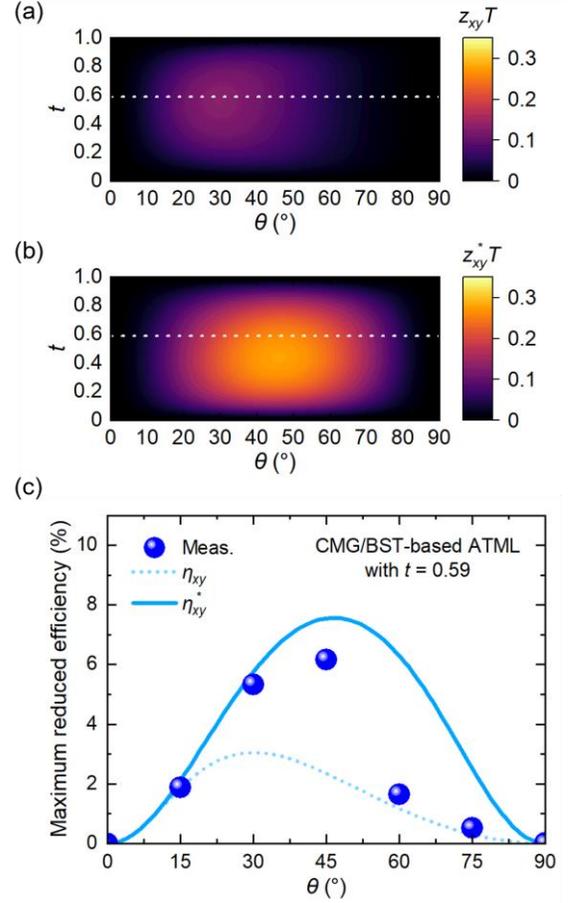

FIG 5. (a)-(b) Contour plots of (a) $z_{xy}T$ and (b) $z^*_{xy}T$ for CMG/BST-based ATML. (c) The $\theta$ dependence of the transverse thermoelectric figure of merit based on the measured $E_x/\nabla_y T$ together with the calculated $z_{xy}T$ and $z^*_{xy}T$ at $t = 0.59$ for CMG/BST-based ATML.

changes from 45° due to the $-\kappa_{xy}S_{xx}/\kappa_{xx}$ term and positions at 64° for CMG/BST-based ATML [Fig 3(c)]. Figure 5(b) shows that, through balancing between high $S^*_{xy}$ and low $\rho^*_{xx}$ and $\kappa^*_{yy}$, $z^*_{xy}T$ reaches 0.28 at $t = 0.44$ and $\theta = 47°$, which provides totally different design parameters from the conventional analytical calculation.

Let us compare the transverse thermoelectric performance between the isothermal and adiabatic conditions. Note that the figure of merit cannot be directly used for the fair comparison because the domains of definition are different as follows:

$$0 < z_{xy}T < \infty, \quad (16)$$

$$0 < z_{xy}^* T < 1. \tag{17}$$

The presence of ODTC and the Seebeck effect makes it difficult to get the relation between $z_{xy}T$ and $z_{xy}^*T$ whereas $z_{xy}^*T = z_{xy}T/(1 + z_{xy}T)$ was obtained under $S_{xx} = \kappa_{xy} = 0$. Then, we introduce the maximum reduced efficiency for transverse thermoelectric conversion in the isothermal and adiabatic limits ($\eta_{xy}$ and $\eta_{xy}^*$), which enables the fair comparison because the thermoelectric conversion efficiency is expressed by $\eta_{\text{Carnot}}\eta_{xy}$ and $\eta_{\text{Carnot}}\eta_{xy}^*$ with the Carnot efficiency $\eta_{\text{Carnot}}$ for both cases:

$$\eta_{xy} = \frac{\sqrt{1 + z_{xy}T} - 1}{\sqrt{1 + z_{xy}T} + 1}, \tag{18}$$

$$\eta_{xy}^* = \frac{1 - \sqrt{1 - z_{xy}^*T}}{1 + \sqrt{1 - z_{xy}^*T}}. \tag{19}$$

By respectively substituting $z_{xy}T = 0.13$ and $z_{xy}^*T = 0.28$ into Eqs. (18) and (19), $\eta_{xy} = 3.1\%$ in the isothermal limit and $\eta_{xy}^* = 8.1\%$ in the adiabatic limit are obtained. Thus, the change rate in the conversion efficiency is estimated to be 165%, which provides a great impact on the transverse thermoelectric performance. Figure 5(c) shows the $\theta$ dependence of $\eta_{xy}$ and $\eta_{xy}^*$ together with that estimated from the measured $E_x/\nabla_y T$ and calculated $\rho_{xx}^*$ and $\kappa_{yy}^*$ for CMG/BST-based ATML with $t = 59°$. The difference in $\eta_{xy}$ and $\eta_{xy}^*$ is due purely to the existence of ODTC and the Seebeck effect. The trend of our measurement result is relatively close to $\eta_{xy}^*$ rather than $\eta_{xy}$, whose maximum value of 6.2% positions at $\theta = 45°$ not $\theta = 30°$. The above calculations and experiments claim that the distinct materials design and performance potential are obtained depending on whether the thermal boundary condition is isothermal or adiabatic.

## IV. DISCUSSION

We discuss what factors influence the thermal boundary condition toward the precise estimation of $\nabla_x T/\nabla_y T$ and the resultant transverse thermoelectric performance. Obviously, the heat dissipation at the side surfaces through the convection and radiation enforces the imperfect adiabatic condition even in our measurement setup. Scudder [26] demonstrated that the $x:y$ aspect ratio of the target material also changes the boundary condition through the thermal short-circuit by heat spreaders, i.e., heat source and sink. The shorter length in the $y$-direction might be one of the reasons why the measured $E_x/\nabla_y T$ for ATMLs in the previous reports were comparable to or less than $S_{xy}$ [33–35,37]. Meanwhile, this work finds that, even if the measurement setup and geometry of the samples were unchanged, $\nabla_x T/\nabla_y T$ and $E_x/\nabla_y T$ gradually deviates from $-\kappa_{xy}/\kappa_{xx}$ with increasing $\theta$ [Figs. 2(g) and 3(d)], meaning that the region of interest changes from the adiabatic to isothermal condition. In light of this observation, the erosion of isothermal boundaries from the heat spreaders ($\nabla_x T = 0$) has a variable influence in this study, i.e., as $\kappa_{yy}$ increases by increasing $\theta$ [Fig. 4(c)], the heat spreaders more and more imposes the isothermal confinement on the positions to measure $\nabla_x T/\nabla_y T$ and $E_x/\nabla_y T$. Thus, $\kappa_{yy}$ also needs to be taken into account to precisely estimate $\nabla_x T/\nabla_y T$ at the region of interest.

*Contact author: ANDO.Fuyuki@nims.go.jp

Finally, we present how to utilize the ODTC-induced performance enhancement in transverse thermoelectric devices. Figure 6 shows a schematic of a lateral thermopile structure as an example. The important point is that two kinds of transverse thermoelectric materials with the same sign of $-\kappa_{xy}/\kappa_{xx}$ and the opposite sign of $S_{xy}^*$, such as our CMG/BT- and CMG/BST-based ATML slabs, are used. By alternately stacking the two materials intermediated by insulator layers and electrically connected side-by-side, a thermally parallel and electrically series circuit is formed. The side surfaces for electrodes need to be thermally isolated from heat spreaders as much as possible. Owing to the unidirectional transverse heat current for each element, the net transverse temperature gradient $\nabla_x T$ will originate without canceling out in the entire module. As a result, the transverse thermopower will be enhanced compared with $S_{xy}$ [Fig. 3(d)] and positively contribute to the output power.

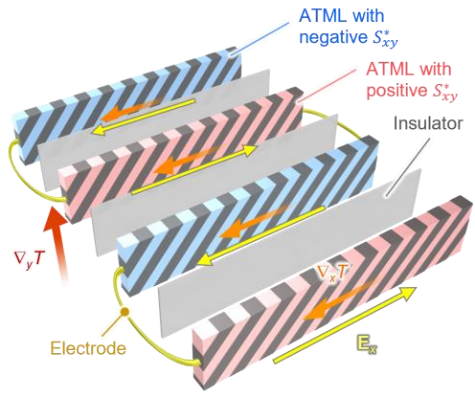

FIG 6. Schematic of a lateral thermopile module utilizing the ODTC-induced transverse thermopower.

## V. CONCLUSIONS

We phenomenologically formulated and experimentally observed an adiabatic transverse thermoelectric performance enhanced by ODTC raising examples of CMG/BT- and CMG/BST-based ATMLs. ODTC induced by an anisotropic thermal conductivity generates a finite transverse temperature gradient and Seebeck-effect-induced thermopower in the adiabatic condition, which is superposed on the isothermal transverse thermopower driven by ODSE. From the two-dimensional temperature distributions on the ATML surfaces in the adiabatic condition, the generation of sizable transverse temperature gradient was observed, which is quantitatively consistent with the calculated ODTC ratio. The resultant adiabatic transverse thermopower in ATMLs was clearly larger than the calculated isothermal ones. By utilizing this ODTC-induced thermopower, we can greatly enhance the transverse thermoelectric performance from the isothermal limit to the adiabatic limit. This work provides a new design for the ODSE materials and devices distinct from the conventional matrix calculations.

## ACKNOWLEDGMENTS

The authors thank H. Sepehri-Amin and A. Takahagi for valuable discussions and K. Suzuki and M. Isomura for technical supports. This work was supported by ERATO "Magnetic Thermal Management Materials" (No. JPMJER2201) from JST, Grants-in-Aid for Scientific Research KAKENHI (No. 24K17610) from JSPS, and NEC Corporation.

## APPENDIX A: CALCURATION OF THERMOELECTRIC TRANSPORT TENSORS FOR ATMLS

First, the thermal conductivity tensor for ATML comprising two materials A and B with different thermal conductivities ($\kappa_A$ and $\kappa_B$) is introduced following Refs. [51–53]. An A/B-based multilayer exhibits the anisotropic thermal conductivities between the directions parallel and perpendicular to the stacking plane as follows:

$$\kappa_\parallel = t\kappa_A + (1-t)\kappa_B, \quad (A1)$$

$$\kappa_\perp = \frac{\kappa_A \cdot \kappa_B}{t\kappa_B + (1-t)\kappa_A}. \quad (A2)$$

When we define $x$- or $z$- and $y$-axes as the parallel and perpendicular directions respectively, the thermal conductivity tensor is expressed using Eqs. (A1) and (A2) as

$$\kappa_{ij} = \begin{pmatrix} \kappa_\parallel & 0 & 0 \\ 0 & \kappa_\perp & 0 \\ 0 & 0 & \kappa_\parallel \end{pmatrix}. \quad (A3)$$

The off-diagonal components are zero in the original description. To transform this plain multilayer into ATML, we introduce a rotation around $z$-axis by an angle $\theta$ which modifies the original axes to $x' = x\cos\theta + y\sin\theta$ and $y' = -x\sin\theta + y\cos\theta$ by the Jacobian matrix of the coordinate transformation:

$$J = \begin{pmatrix} \cos\theta & \sin\theta & 0 \\ -\sin\theta & \cos\theta & 0 \\ 0 & 0 & 1 \end{pmatrix}. \quad (A4)$$

The thermal conductivity tensor is modified through this rotation for the ATML structure, resulting in the finite off-diagonal components as

$$\kappa_{ij} = \frac{J\kappa_{i'j'}J^T}{\det(J)}$$
$$= \begin{pmatrix} \kappa_\parallel \cos^2\theta + \kappa_\perp \sin^2\theta & (\kappa_\parallel - \kappa_\perp)\sin\theta\cos\theta & 0 \\ (\kappa_\parallel - \kappa_\perp)\sin\theta\cos\theta & \kappa_\parallel \sin^2\theta + \kappa_\perp \cos^2\theta & 0 \\ 0 & 0 & \kappa_\parallel \end{pmatrix},$$
$$(A5)$$

where $J^T$ denotes the transpose of $J$, and $\det(J)$ the determinant.

The Seebeck and electrical resistivity tensors ($S_{ij}$ and $\rho_{ij}$) are formulated for A/B-based ATML based on the Goldsmid's method [32]. The Seebeck coefficients and electrical resistivities in the directions parallel and perpendicular to the stacking plane are analytically calculated using thermoelectric transport parameters ($S_{SE,A}$, $S_{SE,B}$, $\rho_A$ and $\rho_B$) as follows:

$$S_{SE,\parallel} = \frac{t\rho_B S_{SE,A} + (1-t)\rho_A S_{SE,B}}{t\rho_B + (1-t)\rho_A}, \quad (A6)$$

$$S_{SE,\perp} = \frac{t\kappa_B S_{SE,A} + (1-t)\kappa_A S_{SE,B}}{t\kappa_B + (1-t)\kappa_A}, \quad (A7)$$

$$\rho_\parallel = \frac{\rho_A \cdot \rho_B}{t\rho_B + (1-t)\rho_A}, \quad (A8)$$

$$\rho_\perp = t\rho_A + (1-t)\rho_B. \quad (A9)$$

*Contact author: ANDO.Fuyuki@nims.go.jp

Following the completely same coordinate transformation by a rotation matrix as Eqs. (A3)−(A5), $S_{ij}$ and $\rho_{ij}$ are obtained. The components relevant to this work are shown as

$$S_{xy} = (S_{\text{SE},\parallel} - S_{\text{SE},\perp}) \sin\theta \cos\theta, \quad (A10)$$

$$S_{xx} = S_{\text{SE},\parallel} \cos^2\theta + S_{\text{SE},\perp} \sin^2\theta, \quad (A11)$$

$$\rho_{xx} = \rho_\parallel \cos^2\theta + \rho_\perp \sin^2\theta. \quad (A12)$$

*Contact author: ANDO.Fuyuki@nims.go.jp

*Contact author: ANDO.Fuyuki@nims.go.jp

*Contact author: ANDO.Fuyuki@nims.go.jp